\documentclass[aps,prl,twocolumn,showpacs,superscriptaddress,floatfix,10pt,amsfonts,amssymb,amsmath]{revtex4-2} 
\usepackage{microtype}
\usepackage[otfmath,varbb,varg]{newtx} 
\usepackage{bm}
\usepackage{graphicx}
\usepackage[dvipsnames,svgnames,x11names,hyperref]{xcolor}
\usepackage{hyperref}
\hypersetup{colorlinks=true, linkcolor=NavyBlue, urlcolor=NavyBlue, citecolor=NavyBlue}
\usepackage{enumitem}
\usepackage[normalem]{ulem}
\usepackage[capitalise]{cleveref}
\usepackage{braket}
\usepackage[cal=euler]{mathalfa}

\newcommand{\la}{\langle}
\newcommand{\ra}{\rangle}
\newcommand{\brho}[1]{\bar{\rho}_{\bm #1}}

\newcommand{\pdagger}{\phantom{\dagger}}
\DeclareMathOperator{\tr}{tr}

\begin{document}
\title{Intraband collective excitations in fractional Chern insulators are dark}
\author{Tobias M. R. Wolf}
\email{tobias.wolf@austin.utexas.edu}
\affiliation{Department of Physics, University of Texas at Austin, Austin, TX 78712, USA}
\author{Yung-Chun Chao}
\affiliation{Department of Electrophysics, National Yang Ming Chiao Tung University, Hsinchu 300, Taiwan}
\author{Allan H. MacDonald}
\affiliation{Department of Physics, University of Texas at Austin, Austin, TX 78712, USA}
\author{Jung Jung Su}
\affiliation{Department of Electrophysics, National Yang Ming Chiao Tung University, Hsinchu 300, Taiwan}
\date{\today}

\begin{abstract}
	The low-energy collective excitations of semiconductors and insulators often couple strongly to light, allowing them to be probed optically.
	We argue here that in fractional Chern insulators intra-band collective excitations are dark in the sense that they couple anomalously weakly to light.
	This conclusion is based on a relationship between ideal quantum geometry and the structure factor of a Chern band, and on a classical plasma analogy motivated by the vortexibility property of ideal Chern bands.
\end{abstract}

\maketitle

\paragraph{Introduction.}
The collective excitations of semiconductors and insulators often couple strongly to light, providing both an experimentally convenient probe for fundamental physics studies and a rich axis for optical and electrooptic applications.
This has been especially true in transition metal dichalcogenide (TMD) two-dimensional semiconductors \cite{wang2018colloquium}.
Recently a novel type of insulating state, the fractional Chern insulator (FCI) \cite{Wen_FCI,KaiSun_FCI,Neupert_FCI,Bernevig_Regnault_FCI,Sheng_FCI}, has been discovered in TMD moiré materials \cite{cai2023signatures,Zeng2023thermodynamic,park2023observation,xu2023fci} and in rhombohedral graphene multilayers \cite{lu2024fractional}.
Fractional Chern insulators are exotic strongly correlated states with fractionalized quasiparticle excitations that have potential applications in topological quantum computing, and hence are of considerable interest.
The appearance of FCI states is thought to be related to weak dispersion of a partially occupied band combined with suitable quantum geometry \cite{roy2014band,Parameswaran_Ideal,ledwith2023vortexability} of that band.
In this paper we argue that the low-energy intraband collective excitations of FCI's are dark in the sense that they couple weakly to light.

\begin{figure*}
	\centering
	\includegraphics[width=1\linewidth]{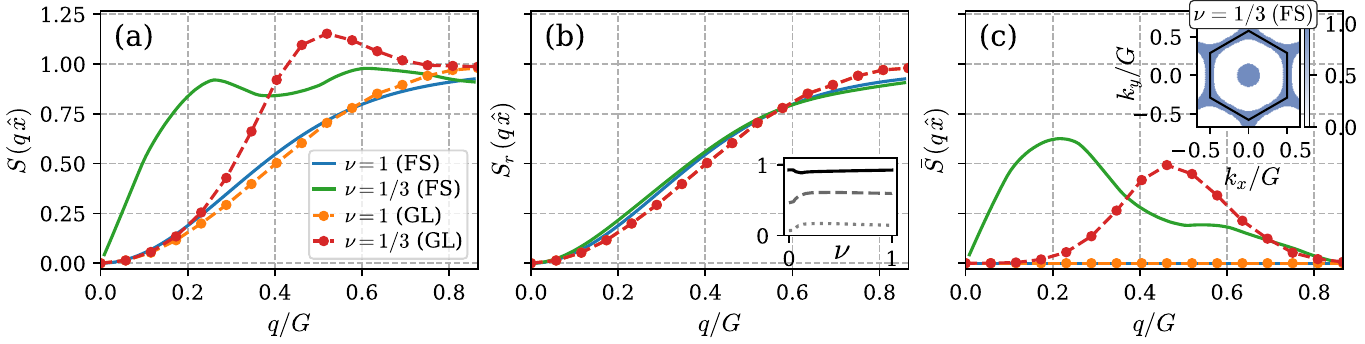}
	\caption{\label{Fig:StaticStructure}
	Total, remote band, and projected static structure factors for generalized Laughlin (GL), and Fermi sea (FS) states.
	(a)~The total structure factor $S(q\,\hat{x})$ at fractional band filling $\nu=1/3$ behaves like $q^2$ at long wavelength in insulating states but like $q$ in Fermi liquid states.
	Corresponding $\nu=1$ structure factors are shown for comparison.
	(b)~The remote band structure factor $S_r(q\,\hat{x})$ of Fermi sea states and generalized Laughlin states at $\nu=1$ and $\nu=1/3$.
	The inset shows the filling factor dependence $S_r(q'\,\hat{x})$ \emph{vs.
	} $\nu$ for the Fermi sea states at select momenta $q'/(\sqrt{3}G/2) = 0.2, 0.5, 1.0 $ (dotted, dashed, and solid lines).
	(c)~The projected static structure factor $\bar{S}(q\,\hat{x})$ vanishes as $q^4$ for generalized Laughlin states but as $q$ for Fermi liquid states.
	The inset shows the momentum-space occupation of the $\nu=1/3$ Fermi liquid state within the first Brillouin zone.
	The particular Fermi sea states here are partially filled flat-bands with near-ideal quantum geometry obtained from the moiré continuum model for homobilayer MoTe$_2$ with reciprocal lattice constant $G=4\pi/(\sqrt{3}a_{\mathrm{M}})$ at twist angle $\theta=3^{\circ}$, see Ref.~\cite{Supplemental} for details.
	}
\end{figure*}

\paragraph{Dynamic structure factor and optical conductivity. } Our conclusions concerning the optical conductivity of fractional Chern insulators follow in part from its formal relationship to a more general quantity, the dynamic structure factor:
\begin{align} \label{eq:dynamicstructure}
	S(\bm{q},E) = \frac{1}{N} \sum_{n} |\braket{ \Psi_n|\rho_{\bm{q}}|\Psi_0}|^2 \, \delta(E-E_n+E_0).
\end{align}
The static structure factor $S(\bm{q})$ can be obtained by integrating $S(\bm{q},E)$, which is positive definite, over excitation energy $E$.
In \cref{eq:dynamicstructure}, $\ket{\Psi_n}$ is a many-body eigenstate, the label $n=0$ is reserved for the ground state, $E_n$ is a many-body energy eigenvalue, and $ \rho_{\bm{q}} = \sum_{i=1}^N \rho_{\bm{q}}^{(i)} = \sum_{i=1}^N \exp(-i\bm{q}\cdot \hat{\bm{r}}_i), $ where $N$ is the total electron number, is the density-operator.

In linear response theory \footnote{ In a system with the discrete translation symmetry of a crystal the density-density response is non-zero when its momentum arguments differ by reciprocal lattice vectors.
	The $\chi$ and $S$ referred to here are diagonal elements of the response matrix.
},
the retarded density-density response function is
\begin{align*}
	\chi(\bm{q},E) =\frac{N}{A} \int_0^{\infty} \!\!\!\!\!\! dE' S(\bm{q},E')
	\Big[ \frac{1}{E-E'+i\eta}-\frac{1}{E+E'+i\eta} \Big],
\end{align*}
such that all density-density response properties of a many-electron system can be expressed in terms
of the dynamic structure factor.  In particular, the object of interest here -- the real part of the
frequency dependent conductivity -- is related to $\chi$ by the
continuity equation, from which it follows that
\begin{align} \label{eq:sigmaandS}
	\sigma^R_{xx}(E) & = \lim_{q\to 0} \frac{-e^2E}{q^2} \chi^I(\bm{q},E) \nonumber \\
	                 & = \frac{N\pi e^2}{A} \lim_{q\to 0}
	\frac{E S(\bm{q},E)}{q^2} .
\end{align}

Our goal is to isolate the contribution to $S(\bm{q},E)$ from low-energy transitions within a partially occupied flat band with label $f$.
For this purpose we define the density operator $\bar{\rho}_{\bm{q}}=\sum_{i=1}^N \bar{\rho}_{\bm{q}}^{(i)}$ projected to flat-band Bloch states $\ket{\psi_{f,\bm{k}}}$ with
\begin{align}
	\bar{\rho}_{\bm{q}}^{(i)}= \sum_{\bm{k}\in\mathrm{BZ}} \;
	|\psi_{f,\bm{k}-\bm{q}}\ra\la \psi_{f,\bm{k}-\bm{q}}| e^{-i\bm{q}\cdot \hat{\bm{r}}_i}
	|\psi_{f,\bm{k}}\ra\la \psi_{f,\bm{k}}|,
	\label{eq:rhobar}
\end{align}
and a corresponding projected dynamic structure factor,
\begin{align}
	\bar{S}(\bm{q},E) = \frac{1}{N} \sum_{n} |\langle \Psi_n|\bar{\rho}_{\bm{q}}|\Psi_0\rangle|^2 \, \delta(E-E_n+E_0).
	\label{eq:dynamicstructure_proj}
\end{align}
We now argue that the contribution to $\sigma^R_{xx}(E)$ from $\bar{S}(\bm{q},E)$ is very small in FCI states.
Our argument is based on a relationship we establish between $\bar{S}(\bm{q})$ and quantum geometry, and on the properties of incompressible short-range-interaction ground states in vortexible bands.

\paragraph{Flat and remote band contributions to $S(\bm{q})$.}
We work to leading order in the ratio of interactions to the energy separation between flat and remote bands, and assume as an inessential convenience that the flat band is lowest in energy.
In this limit many-body states can be classified by the number of electrons in the flat bands and a low-energy sector can be identified in which all electrons occupy the flat band.
The projected static structure factor provides a measure of the strength of the density-response within this sector, and the coefficient of $q^2$ in its long-wavelength limit a measure of its contribution to the optical conductivity.
To separate the
projected contribution from $S(\bm{q})$, we note that
\begin{align}
	S(\bm{q}) & \equiv \frac{1}{N} \braket{\rho_{-\bm{q}} \rho_{\bm{q}}}_0 = \frac{1}{N} \sum_{ij} \braket{e^{i\bm{q}\cdot\hat{\bm{r}}_i} e^{-i \bm{q}\cdot\hat{\bm{r}}_j}}_0 \end{align} can be separated into same ($i=j$) and distinct ($i\neq j$) particle contributions, and that band-projection is relevant only for the former \footnote{ See \cref{eq:bloch_innerproduct} in the Supplemental Material~\cite{Supplemental} for details on the first Brillouin zone mapping of inner products.
}:
\begin{align*}
	{S}_{ii} (\bm q)
	= \frac{1}{N}
	\sum_{b\bm k}
	\braket{u_{f, \bm k} | u_{b, \bm k+\bm q} }
	\braket{u_{b, \bm k+\bm q} | u_{f, \bm k} }
	n_{f,\bm{k}} ,
\end{align*}
where $b$ labels band indices, $|u_{b, \bm k}\ra=e^{-i\bm{k}\cdot\hat{\bm{r}}}|\psi_{b,\bm{k}}\ra$ is the periodic part of the Bloch state, and $n_{f,\bm{k}}=\braket{a^{\dagger}_{f, \bm k} a^{}_{f, \bm k}}_0$ is the occupation number.
The projected static structure differs only in that band summation is restricted to $b=f$, such that
\begin{align} \label{eq:remoteband}
	S_r(\bm{q}) & = S(\bm{q})-\bar{S}(\bm{q}) \\ & = \sum_{\bm k} \Big[ 1 - |\braket{ u_{\bm k+\bm q} | u_{\bm k} }|^2 \Big] \, n_{\bm{k}} \nonumber\end{align} is the remote band contribution to the static structure factor, where quantities without band label are implicitly assumed to be in the flat band.

At long-wavelengths, we thus find that the remote band  structure factor $S_r(\bm{q})$ is the integrated Fubini-Study tensor weighted by band occupation,
\begin{align} \label{eq:remoteband_g}
	\lim_{q \to 0}
	S_r(\bm{q}) & = q_{\mu} q_{\nu} \sum_{\bm k} g_{\mu \nu}^f(\bm k) \, n_{f,\bm{k}},\end{align} where $|\la u_{b, \bm k-d\bm q} | u_{b, \bm k} \ra|^2 = 1 - g_{\mu \nu}^b(\bm k) \,dq_{\mu} dq_{\nu}$ defines the Fubini-Study tensor $g_{\mu \nu}^b$ \cite{roy2014band} with implied sum over $\mu,\nu$.
If, as we shall argue below,
(A)~the ground state band-occupation numbers $\braket{ \Psi_0 | a^{\dagger}_{f, \bm k} a^{}_{f, \bm k} | \Psi_0 }$
are $\bm{k}$-independent,
(B)~the ideal quantum geometry trace condition $\tr [g^f (\bm k)] = {\Omega^f (\bm k)}$
holds, where $\Omega^f (\bm k)$ is the Berry curvature,
(C)~and the system has sufficient symmetries such that integrated values of $g^f_{xx}$ and $g^f_{yy}$ are identical, then
\cref{eq:remoteband_g} leads to
\begin{align} \label{eq:remoteband_g_ideal}
	\lim_{q \to 0} \frac{S_r(\bm{q})}{q^2} = C^f \, \frac{A_0}{4\pi},
\end{align}
where $C^f$ is the Chern number of band $f$.

We note from \cref{eq:remoteband_g} that $S_r(\bm{q})$ vanishes quadratically in $q$ for any many-body state, including FCI and Fermi liquid states, as illustrated in \cref{Fig:StaticStructure}(b).
In contrast, the full static structure factor $S(\bm{q})$ has very different properties in these two cases: for FCI states, it vanishes quadratically in $q$, while for Fermi liquid states, it vanishes linearly because of the singularities introduced by the Fermi surface, see \cref{Fig:StaticStructure}(a).
We now show that the coefficents of $q^2$ for FCI states are identical in the remote band and full structure factors, $S_r(\bm{q})$ and $S(\bm{q})$.

\paragraph{Plasma analogy, perfect screening, and FCI state static structure factors.}
Our analysis is based on the premise that FCI states appear when Chern bands have (near-)ideal quantum geometry and vortexable bands \cite{ledwith2023vortexability} and therefore that the incompressible ground states for short-range interactions \cite{roy2014band,wang2021exact,ledwith2023vortexability} are generalized Laughlin many-body wavefunctions \footnote{ While not yet established experimentally, there are strong arguments \cite{morales2023pressure,morales2023magic} that this connection to quantum geometry applies to all moir\'e material FCI states.
},
\begin{align} \label{eq:Laughlin}
	\Psi_m(\bm{r}_1,\dots,\bm{r}_N) = \prod_{i<j}^{N}(z_i\!-\!z_j)^m\prod_{k=1}^{N}e^{-\frac{|z_k|^2}{4\ell^2}-\Phi(\bm{r}_k)}.
\end{align}
In \cref{eq:Laughlin}, $m$ is an odd integer, $z_k=x_k+iy_k$ is position expressed as a complex number, $\ell$ is the magnetic length corresponding to a quantum of pseudo magnetic field \cite{morales2023magic} per unit cell area $A_0$ (i.e., $2\pi\ell^2=A_0$), and $\Phi(\bm{r})$ is periodic function with zero average value that distinguishes one flat band from another.

The wavefunction $\Psi_m$ is a product of polynomial and Gaussian factors that respectively increase and decrease in magnitude exponentially with $N$; the factors must balance in order for the $|\Psi_m|^2$ distribution function to have a sensible thermodynamic limit, implying $m N =N_{\phi}$, where $N_{\phi}=A/A_0$ is the number of periods of the moir\'e potential in the system.
To make this argument, we can view ${|\Psi_m|^2\propto\exp(- U(\bm{r}_1,\dots,\bm{r}_N)/m)}$ as the classical Boltzmann weight of a 2D Coulomb plasma \cite{Laughlin} containing particles with charge $q_i=m$.
With this identification, \cref{eq:Laughlin} leads to
\begin{align}
	U = - \sum_{i=1}^N q_i \phi_{\text{b}}(\bm{r}_i) + \sum_{i<j}^{N} q_i V(\bm{r}_i-\bm{r}_j) q_j ,
	\label{eq:plasma}
\end{align}
where $V(\bm{r})\equiv-2\ln|\bm{r}|$ and
\begin{align}
	\phi_{\text{b}}(\bm{r})= \int d^2r' \, V(\bm{r}-\bm{r}') \, \rho_{\text{b}}(\bm{r}')
\end{align}
is the attractive 2D Coulomb potential produced by the background charge
$\rho_{\text{b}}(\bm{r})= (2\pi\ell^2)^{-1} + \nabla^2 \Phi(\bm{r})/2\pi \equiv A^{-1}\cdot \sum_{\bm{G}} e^{i\bm{G}\cdot\bm{r}}\rho_{\text{b}\bm{G}}$ with $\rho_{\text{b}\bm{0}}=A/(2\pi\ell^2)$ \footnote{Since $\nabla^2 u(\bm{r})$ is periodic, it does not contribute to the total charge density.}.
The plasma distribution function has a sensible thermodynamic limit only when it is charge neutral ($\rho_{\text{b}\bm{0}}=m\rho_{\bm{0}}=mN$), capturing the constraint $mN=N_{\phi}$.
The average particle density then is $\bar{n}=N/A=(m 2\pi\ell^2)^{-1}=1/(mA_0)$, implying that the band filling factor for this state is $\nu=1/m$.
For the calculations we describe below it is convenient to write $U$ in momentum space using $\rho_{\bm{k}}=\sum_{i} e^{-\bm{k}\cdot\bm{r}_i}$, such that
\begin{align} \label{eq:plasmamomentum}
	U = \frac{1}{2A} \sum_{\bm{k}\ne 0} \frac{4\pi }{|\bm{k}|^2} \; | m \rho_{\bm{k}}- \rho_{\text{b}\bm{k}}|^2,\end{align} up to constant terms that do not change the relative weight.

Next we argue that the average density in any region much larger than one unit cell must approach $\nu/A_0$ even when one particle is held fixed near the center of that region.
In the plasma analogy this property is known as the Stillinger-Lovett \cite{stillinger1968general,stillinger1968ion} perfect screening condition, which implies that the plasma
inverse dielectric function vanishes in the long-wavelength limit:
\begin{align}
	\epsilon^{-1}(\bm{q}) = 1 + \frac{4\pi m}{q^2} \chi(\bm{q}) \underset{q\to 0}{\longrightarrow} 0 .
	\label{eq:stillov}
\end{align}
Since the plasma response function $\chi(\bm{q})$ and $S(\bm{q})$ are both Fourier
transforms of the two-particle density correlation function (with different normalization),
\cref{eq:stillov} implies
\begin{align}\label{eq:structurefactor_perfectscreening}
	\lim_{q \to 0} \frac{S(\bm{q})}{q^2} = \frac{\ell^2}{2}=\frac{A_0}{4\pi}.
\end{align}
Given the assumption (see proof below) that \cref{eq:remoteband_g_ideal} applies to generalized Laughlin states, \cref{eq:structurefactor_perfectscreening} implies that the dipole contribution to $S(\bm{q})$ is exhausted by the remote-band response, leaving no room for flat-band contributions.

To show that \cref{eq:remoteband_g_ideal} applies, we demonstrate that the band occupation probabilities for generalized Laughlin states are $\bm{k}$-independent not just for ${m=1}$ (${\hat{n}_{f,\bm{k}}=1}$), but for any fractional filling ${m>1}$.
The argument requires two ingredients:
(i)~band vortexibility~\cite{ledwith2023vortexability} together with the density matrix properties in Ref.~\cite{macdonald1988density} non-trivially implies that two vortexable band states with proportional charge densities have proportional one-particle density matrices \cite{macdonald1988density}, and thus proportional occupation numbers according to
\begin{align}\label{eq:occupation_densitymatrix}
	\braket{\hat{n}_{f,\bm{k}}}
	 & = \int \!\!\! d^2r d^2r'\, \psi_{f,\bm{k}}^*(\bm{r}) \, \rho^{(1)}(\bm r,{\bm r}') \, \psi_{f,\bm{k}}(\bm{r}'),
\end{align}
and, (ii)~that the charge density of a generalized Laughlin state (${m>1}$) is approximately ${1/m}$ times the charge density of a filled flat band (${m=1}$).

\begin{figure}
	\centering
	\includegraphics[width=1\linewidth]{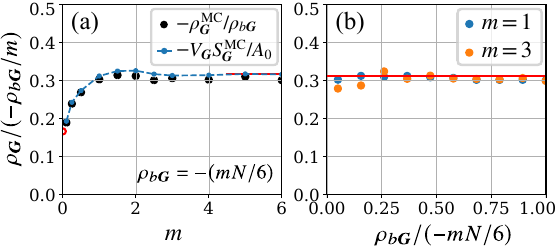}
	\caption{Dependence of the first reciprocal lattice vector shell charge density modulation $\rho_{\bm{G}}$
		on the plasma coupling parameter $m$ and the background modulation $\rho_{\text{b}\bm{G}}$.
		(a) Dependence of the $\rho_{\bm{G}}$ calculated by sampling the plasma distribution function on the plasma coupling parameter $m$.
		The exact charge modulation (black) is well-described by the linear response relation (blue) when Monte-Carlo results are used for the unperturbed system structure factor, and agrees with analytical results in the limiting cases $m\to 0$ and $m\gg 1$ (red).
		(b) Dependence of $\rho_{\bm{G}}$ on background density modulation $\rho_{\text{b}\bm{G}}$ a$m=1$ and $m=3$, showing that all considered background modulations are well within the linear response regime.
	}
	\label{fig:plasma}
\end{figure}

To establish property (ii), we evaluate and investigate the $m$-dependence of the Fourier components of the charge density $\langle \rho_{\bm{k}}\rangle=\langle \Psi_m | \rho_{\bm{k}} |\Psi_m\rangle$ by Monte-Carlo sampling of the plasma analogy Boltzmann distribution in \cref{eq:plasmamomentum}, see~\cite{Supplemental} for details.
In linear response~\cite{Supplemental}, the induced charge density at wavevector $\bm{G}$ is proportional
to the background charge density at that wavevector and to the static structure factor of the unperturbed uniform system:
\begin{align} \label{eq:linearresponse}
	\langle\rho_{\bm{G}}\rangle = - \frac{1}{A_0} \frac{4\pi }{|\bm{G}|^2}
	S(\bm{G}) \bigg|_{\rho_{\text{b}\bm{G}}=0} \; \frac{\rho_{\text{b}\bm{G}}}{m}.
\end{align}
To see how \cref{eq:linearresponse} justifies property (ii), we note that the electron density ${\bar{n}=1/(mA_0)}$ sets the natural length scale for the structure factor ${S(\bm{q})}$: For the first shell, $|\bm{G}|\, \bar{n}^{-1/2} = (8\pi^2m/\sqrt{3})^{1/2} \approx 6.75 \sqrt{m}$ is large already at ${m=1}$ which leads to uncorrelated values ${S(\bm{G})\approx 1}$.
For $m\geq 1$, we thus estimate $\langle\rho_{\bm{G}}\rangle \approx -(4\pi/{A_0}|\bm{G}|^2) (\rho_{\text{b}\bm{G}}/m) = -\sqrt{3}/(2\pi) (\rho_{\text{b}\bm{G}}/m)\approx -0.05 N$.
In \cref{fig:plasma}, we show the first-shell charge density $\rho_{\bm{G}}$ {\it vs.
	} $m$ in the presence of a non-uniform background charge, obtained using Monte Carlo to sample the plasma distribution \cite{Supplemental}.
We find that calculating $\rho_{\bm{G}}$ by sampling the non-uniform system gives the same results as \cref{eq:linearresponse} with the uniform system correlation function.
In the regime of positive definite background charge density studied here, our calculations verify the accuracy of \cref{eq:linearresponse}.
We further remark that for ${m \to 0}$, $|\bm{G}|\, \bar{n}^{-1/2}$ being small justifies the perfect screening limit of \cref{eq:structurefactor_perfectscreening} and yields $m\rho_{\bm{G}} = -\rho_{\text{b}\bm{G}}$, in agreement with direct Monte-Carlo evaluation, see \cref{fig:plasma}.
For ${m\to 0}$, the charge density thus locally cancels the background density, and becomes insensitive to plasma interactions.
On physical grounds, we expect that our result for both limits survive into the non-linear regime of more strongly varying charge densities.
Importantly for this work, we found property~(ii): at fractional filling $\nu=1/m$ the charge density [by (i) also the band occupaition] is simply $1/m$ times that of the filled-band.

\begin{figure}[t]
	\centering
	\includegraphics[width=.8\linewidth]{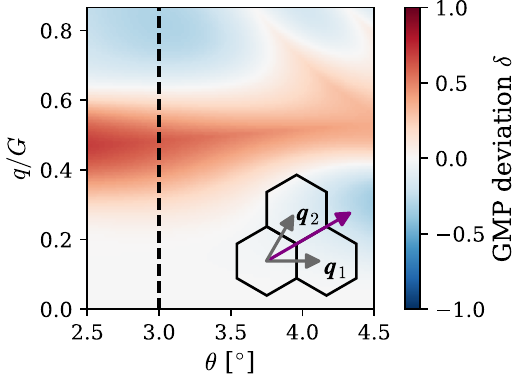}
	\caption{
		GMP algebra deviation $\delta(q)$ of the projected structure factor in the nearly-flat moiré valence band of homobilayer MoTe$_2$ as function of twist angle $\theta$.
		We quantify the violation with $\delta(q) \equiv |\la \psi_{f,\bm k+\bm q_1+\bm q_2} |\left[ \rho_{\bm q_1}, \rho_{\bm q_2} \right] | \psi_{f,\bm k} \ra| - |\la \psi_{f,\bm k+\bm q_1+\bm q_2} | 2i \sin(\bm q_1 \wedge \bm q_2 \cdot \bm \Omega(\bm k)/2) \, \rho_{\bm q_1+\bm q_2} \,| \psi_{f,\bm k} \ra| $ with $\bm k=0$ and $\bm q_1, \bm q_2$ as indicated in the inset (${|\bm q_1|=|\bm q_2|=q}$).
		The highlighted angle ${\theta =3{}^{\circ}}$ has both smallest flatband width and smallest trace condition deviation.
		See Ref.~\cite{Supplemental} for details on the continuum model.
	}
	\label{fig3}
\end{figure}

\paragraph{Discussion.}
We have so far shown that dipole transitions within an ideal band are weak.
This property of ideal Chern bands is shared with Landau bands, in which dipole transitions are entirely inter Landau level by virture of Kohn's theorem \cite{kohn1961cyclotron}.
It is natural to examine the degree to which the two cases are similar at larger momentum.
The properties of density response functions within a Landau level are constrained by the GMP algebra \cite{girvin1986magneto} of projected density operators.
\begin{align} \label{eq:gmp_algebra}
	 & [\bar{\rho}_{\bm q_1}, \brho{q_2}]  =
	2 i \, \sin (\bm q_1 \!\wedge\! \bm q_2 \ell^2/2)  \,
	e^{\bm q_1 \cdot \bm q_2 \ell^2/2} \; \bar{\rho}_{\bm q_1+\bm q_2}.
\end{align}
The corresponding operator projected to a Chern band maps Brillouin zone momentum $\bm k$ to
$\bm k+\bm q_1+ \bm q_2$ like the Landau level case, but with
a coefficient that depends on the starting $\bm k$:
\begin{align}
	 & \la \psi_{\bm k+\bm q_1 + \bm q_2} |  [\bar{\rho}_{\bm q_1}, \brho{q_2}] | \psi_{\bm k} \ra = \\
	 &
	\quad \la u_{\bm k+\bm q_1 + \bm q_2}|
	u_{\bm k+ \bm q_2} \ra \la u_{\bm k+\bm q_2} | u_{\bm k}\ra
	-
	\la u_{\bm k+\bm q_1 + \bm q_2}|
	u_{\bm k+ \bm q_1} \ra \la u_{\bm k+ \bm q_1} | u_{\bm k} \ra, \nonumber
\end{align}
where we defined $\ket{u_{\bm{k}'b}}=e^{-i\bm{g}\cdot\hat{\bm{r}}}\ket{u_{\lfloor\bm{k}'\rfloor b}}$ with momenta ${\lfloor\bm{k}'\rfloor=\bm{k}'-\bm{g}}$ reduced to the first Brillouin zone, see Ref.~\cite{Supplemental} for details.
The deviation of projected density operator matrix elements from GMP algebra values is plotted in Fig.~\ref{fig3} as a function of $|\bm q_1|$ and twist angle for $|\bm q_1|=|\bm q_2|$ and a $60^\circ$ angle between the two wavevectors.
These calculations are for the same model of MoTe$_2$ homobilayer moirés used to produce the Fermi liquid structure factors in Fig.~\ref{Fig:StaticStructure}, where we used the twist angle $\theta \approx 3^{\circ}$ at which ideal quantum geometry is most closely approached.
We see in Fig.~\ref{fig3} that the GMP algebra is approximately satisfied for small $\bm q$ over a wide range of twist angles surrounding the most ideal one, and that the agreement with GMP algebra is poor at large $\bm q$.
These results suggest that optical matrix elements will have a strong tendency to be suppressed in Chern bands even if they are relatively far from ideal - and perhaps that the interaction physics responsible for the fractional quantum anomalous Hall effect, which is most easily explained when bands are prefectly ideal, is also tolerant to deviations.

Although dipole allowed transitions are exceptionally weak for intra-band excitations, the oscillator strengths are not precisely zero, there is some hope that optical probes will still prove useful.
We do expect oscillator strengths to increase with disorder and deviations from ideality in the Chern bands just as, in the non-anomalous strong-mangetic-field fractional quantum Hall effect case, disorder-induced Landau band mixing provides dipole access \cite{macdonald1988density,antoniou1992magnetoplasmons} to intra-Landau level transitions via inelastic light scattering \cite{pinczuk1988observation} to short wavelength excitations.
The conclusions of this paper are intended to apply rather broadly to fractional Chern insulator states, although we have performed specific calculations for a model of TMD AA moiré homobilayers \cite{wu2019topological}, the system in which FCI states have recently been observed.
In real systems like this, ideal quantum geometry will never be perfectly realized.
Nearly ideal quantum geometry is however apparently sufficient for Landau-level-like correlations at long length scales, as characterized for example by the density-matrix properties summarized in Fig.~\ref{fig3}.

\begin{acknowledgments}
	\paragraph{Acknowledgments.}
	This work was supported by the Simons Foundation and the Welch Foundation.
	Y.-C.C and J.-J.S acknowledge the support from Ministry of Science and Technology, Taiwan 112-2112-M-A49-039, 110-2124-M-A49 -008 -MY3 and Center for Theoretical and Computational Physics (CTCP) of NYCU.
\end{acknowledgments}

\bibliography{references}

\clearpage
\onecolumngrid

\setcounter{secnumdepth}{2}
\setcounter{equation}{0}
\setcounter{figure}{0}
\setcounter{table}{0}
\makeatletter
\renewcommand{\theequation}{S\arabic{equation}}
\renewcommand{\thefigure}{S\arabic{figure}}
\renewcommand{\thetable}{S\arabic{table}}
\renewcommand{\bibnumfmt}[1]{[S##1]}

\setlength{\parindent}{0pt}
\setlength{\parskip}{6pt plus 2pt minus 1pt}

\section*{Supplementary Material}

\section{Moiré bands of AA-stacked transition metal dichalcogenide moir\'e  homobilayers}

\paragraph{Model} Following Ref.~\cite{wu2019topological}, we describe the effective electronic bandstructure of AA-stacked transition metal dichalcogenide homobilayers using the moiré Hamiltonian
\begin{align}\label{eq:moireHamiltoinan_WuMacDonald}
	H_{K\uparrow}(\bm{k})=  \begin{pmatrix}
		                        -\frac{\hbar^2(\bm k - \bm{\kappa}_{+})^2}{2m^*} + \Delta_b (\bm r) & \Delta_T (\bm r)                                                    \\
		                        \Delta_T^{\dagger} (\bm r)                                          & -\frac{\hbar^2(\bm k - \bm{\kappa}_{-})^2}{2m^*} + \Delta_t (\bm r)
	                        \end{pmatrix},
\end{align}
where $\bm{k}$ is the Bloch momentum, $m^*$ is the effective valence band mass at valley $K$ and spin $\uparrow$, where $K$ and $K'$ are the inequivalent corners of the lattice-scale Brillouine zone.
The momenta $\bm{\kappa}_{\pm}$ are the two inequivalent corners of the moiré Brillouin zone constructed from the first-shell moiré wavevectors $\bm{G}_j=G\,(\cos(j 2\pi/6),\sin(j 2\pi/6))$, where $G=4\pi/(\sqrt{3}a_{\mathrm{M}})$.
The moiré modulation in \cref{eq:moireHamiltoinan_WuMacDonald} is $\Delta_{b,t}(\bm r) = 2V \sum_{j=1,3,5} \cos(\bm G_j \cdot \bm{r} \pm \psi)$ and $\Delta_{T} (\bm r) = w\,(1+ \exp(-i \bm G_2 \cdot \bm r) +\exp(-i \bm G_3 \cdot \bm r))$.
Here we use $a_{\mathrm{M}}=0.352$~nm, $\left(m^*, v, \psi, w\right)=\left(0.6 m_{\mathrm{e}}, 20.8 \text{ meV},+107.7^{\circ},-23.82 \text{ meV}\right)$ for MoTe$_2$ from Refs.~\cite{xu2023fci,Wang2024LSDFT}.
When applicable, the moiré Hamiltonian $H_{K'\downarrow}(\bm{k})$ is determined by time-reversal symmetry.

\begin{figure}[b]
	\centering
	\includegraphics[width=1.\linewidth]{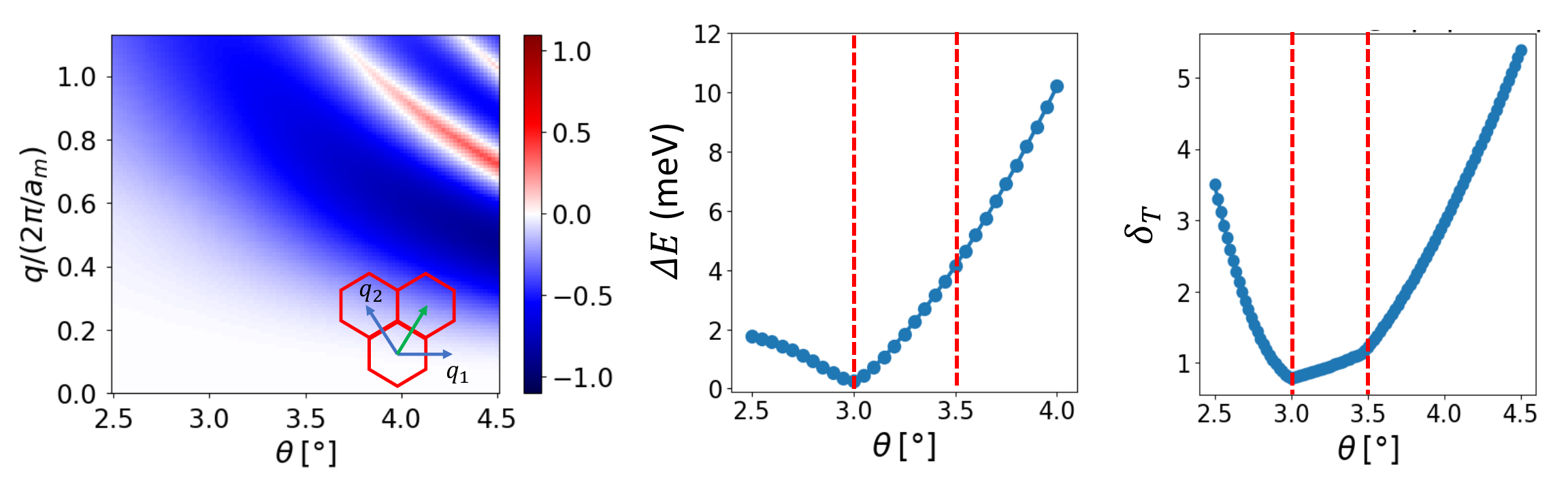}
	\caption{Dependence of the quantum geometry and flatness of the moiré valence band of homobilayer MoTe$_2$ on the twist angle $\theta$. (a)~GMP algebra deviation $\delta(q)$, where inset indices the chosen linecut of the more generial higher-dimensional GMP relation. (b)~Bandwidth highlighting the magic angle. (c)~Deviation from the ideal trace condition, highlighing an optimal range. }
	\label{figS1}
\end{figure}

\paragraph{Magic angle and ideal quantum geometry.}
If a given Chern band's states can be mapped to a lowest Landau level, the formation of interacting FCI ground states becomes a plausible consequence of the Landau level physics.
Various metrics have been proposed to assess if and to what extent such a representation is plausible, including (1) how closely the flat-band projected density operators satisfy the Girvin-MacDonald-Platzman (GMP) algebra, and (2) if the bands are vortexable according to a trace condition.
Both measures reflect the band's quantum geometry.
To measure how much the projected density operators deviate from satisfying the GMP algebra, we consider $\delta(q) \equiv |\la \psi_{\bm k+\bm q_1+\bm q_2} |\left[ \rho_{\bm q_1}, \rho_{\bm q_2} \right] | \psi_{\bm k} \ra | - |\la \psi_{\bm k+\bm q_1+\bm q_2} | 2i \sin(\bm q_1 \wedge \bm q_2 \cdot \bm \Omega(\bm k)/2) \, \rho_{\bm q_1+\bm q_2} \,| \psi_{\bm k} \ra | $.
For the moiré flatband considered in this work, the quantum geomety is nonuniform and the GMP deviation $\delta$ depends on the direction: In Fig.~3 of the main text, we show the GMP deviation $\delta(q)$ for $(\bm q_1, \bm q_2)$ in the direction of $(0, 60 ^{\circ})$.
In the left panel of Fig.
\ref{figS1}, we show the GMP deviation for $(\bm q_1, \bm q_2)$ in the direction of $(0, 120 ^{\circ})$.
For these two cases of $(\bm q_1, \bm q_2)$, the deviation always increases with increasing twist angle (in the shown range).
A potentially more general measure is how much the quantum geomtry deviates form the vortexibility trace condition, meaning $ \delta^{T} \equiv \sum_{\bm k \in \mathrm{BZ}} \tr [g^f (\bm k)] - {\Omega^f (\bm k)}.
$
The middle panel of \cref{figS1} shows the dependence of the bandwidth $\Delta E$ of the flatband on twist angle $\theta$ and the right panel shows the deviation from trace condition $\delta^{T}$.
In the main text, we focus on $\theta = 3^{\circ}$ because it optimizes both the bandwidth and the trace condition.

\section{Particle density and structure factor}

In the $N$-particle product-space representation, the particle density operator in momentum space is
\begin{align}
	\hat{\rho}_{\bm{q}} = \int d^dr\, e^{-i\bm{q}\cdot\bm{r}} \hat{\rho}(\bm{r}) = \sum_{i=1}^N \hat{\rho}_{\bm{q}}^{(i)} = \sum_{i=1}^N e^{-i\bm{q}\cdot \hat{\bm{r}}_i},
\end{align}
and the two-particle density operator in momentum space is
\begin{align}
	\hat{S}_{\bm{q}} = \frac{1}{N} \hat\rho_{-\bm q} \hat\rho_{\bm q}
	= \frac{1}{N}  \sum_{i,j=1}^N \hat{\rho}_{-\bm{q}}^{(i)} \hat{\rho}_{\bm{q}}^{(j)}
	= \hat{\rho}_{\bm{q}=0} \, + \frac{1}{N} \sum_{i\neq j} e^{i\bm{q}\cdot \hat{\bm{r}}_i} e^{-i\bm{q}\cdot \hat{\bm{r}}_j},
\end{align}
where we see that the operator can be decomposed into single-particle and two-particle contributions.

Physically, we are interested in the susceptibility and, equivalently, the dynamical structure factor,
\begin{align}
	\chi(\bm{q}, E)
	             & =\frac{1}{A} \sum_{m, n} \frac{e^{-\beta E_m}}{Z}\left\{\frac{\left|\braket{\Psi_m|\hat{\rho}_{\bm{q}}| \Psi_n}\right|^2}{E-E_m+E_n+i \eta}-\frac{\left|\braket{\Psi_m|\hat{\rho}_{\bm{q}}| \Psi_n}\right|^2}{E+E_m-E_n+i \eta}\right\}
	=\frac{N}{A}\int_{-\infty}^{\infty} \!\!\!\! d E' S(\bm{q},E')\left\{\frac{1}{E+E'+i \eta}-\frac{1}{E-E'+i \eta}\right\},
	\nonumber                                                                                                                                                                                                                                              \\
	S(\bm{q}, E) & =\frac{2 \pi}{N Z} \sum_{m, n} e^{-\beta \hbar E_m}\left|\braket{\Psi_m|\hat{\rho}_{\bm{q}}| \Psi_n}\right|^2 \delta\left(E-E_n+E_m\right),\end{align} where $\beta$ is the inverse temperature, $Z$ is the partition function.
From these definitions, it is straightforward to show that
\begin{align}
	S(\bm{q})\equiv \int_{-\infty}^{\infty} \frac{d E'}{2 \pi} S(\bm{q}, E') & = \frac{1}{N}\braket{\hat{\rho}^{\dagger}_{\bm{q}} \hat{\rho}_{\bm{q}}} = \braket{\hat{S}_{\bm{q}}}, \nonumber \\ \int_{-\infty}^{\infty} \frac{d E'}{2 \pi} E' S(\bm{q}, E') & = \frac{1}{N} \braket{\hat{\rho}_{\bm{q}}\left[\mathcal{H}, \hat{\rho}_{\bm{q}}^{\dagger}\right]} =\frac{1}{2N}\braket{\left[\hat{\rho}_{\bm{q}},\left[\mathcal{H}, \hat{\rho}_{\bm{q}}^{\dagger}\right]\right]}=\frac{\bm{q}^2}{2m}, \end{align} where $\mathcal{H}$ is the many-body Hamiltonian.
The relation in the second line is the $f$-sum rule for the dynamic structure factor, and the last equality follows for the specific case $\mathcal{H}=-\sum_{i=1}^N \bm{\nabla}_i^2/(2 m)+V\left(\bm{r}_1, \ldots, \bm{r}_N\right)$.

The electromagnetic response for charge density $\rho$ and charge current $\bm{j}$ in presence of external fields is
\begin{align}
	 & \rho(\bm{q}, \omega)=-\chi_{\rho}(\bm{q}, \omega) V_{\mathrm{ext}}(\bm{q}, \omega), \\
	 & \bm{j}(\bm{q}, \omega)=\sigma(\bm{q}, \omega) \bm{E}(\bm{q}, \omega),
\end{align}
where $\bm{E}(\bm{q}, \omega)=-i \bm{q}
	V_{\mathrm{ext}}(\bm{q}, \omega)$ is the electric field induced by an external potential $V_{\mathrm{ext}}(\bm{q}, \omega)$.
As stated in the main text, the continuity equation $\partial \rho / \partial t+\nabla \cdot \bm{j}=0$ relates these two responses:
\begin{align}
	\chi_{\rho}(\bm{q}, \omega)=i\,  q_\alpha q_\beta \frac{ \sigma_{\alpha\beta}(\bm{q}, \omega)}{\omega} .
\end{align}
This leads to \cref{eq:sigmaandS}, if we further replace charge density-response with a particle-density response, i.e., $\chi_{\rho} = e^2 \chi$.

\subsection{Band projected density and structure factor}

To study band-projected quantities, we introduce the projection $\mathcal{P}_b=\prod_i \mathcal{P}^{(i)}_b=\mathcal{P}_b^{\dagger}=\mathcal{P}_b^2$ onto band $b$ and define the \emph{band-projected} density and structure factor
\begin{align}
	\hat{\rho}_{\bm{q}b}=\mathcal{P}_b\hat{\rho}_{\bm{q}}\mathcal{P}_b,
	\qquad
	\hat{S}_{\bm{q}b}
	=\mathcal{P}_b\hat{S}_{\bm{q}}\mathcal{P}_b
	=\frac{1}{N}\mathcal{P}_b\hat\rho_{-\bm q} \hat\rho_{\bm q}\mathcal{P}_b.
\end{align}
The structure factor $\hat{S}_{\bm{q}b}$ should not be confused with the \emph{projected-density} structure factor
\begin{align}
	\hat{\mathcal{S}}_{\bm{q}b}
	=\frac{1}{N}\hat{\rho}_{-\bm{q}b}\hat{\rho}_{\bm{q}b}
	=\frac{1}{N}\mathcal{P}_b\hat\rho_{-\bm q}\mathcal{P}_b\hat\rho_{\bm q}\mathcal{P}_b,
\end{align}
which contains an additional projection to ensure that we never leave the subspace of band $b$.
The difference between these two structure factors is solely from same-particle density-density-correlations:
\begin{align} \label{eq:S_diff}
	\hat{S}^r_{\bm{q}b} = \hat{S}_{\bm{q}b}-\hat{\mathcal{S}}_{\bm{q}b}
	= \frac{1}{N} \mathcal{P}_b\hat\rho_{-\bm q}(1-\mathcal{P}_b)\hat\rho_{\bm q}\mathcal{P}_b
	= \frac{1}{N} \mathcal{P}_b\sum_{ij} \hat\rho^{(i)}_{-\bm q}(1-\mathcal{P}_b)\hat\rho^{(j)}_{\bm q} \, \mathcal{P}_b
	= \frac{1}{N} \mathcal{P}_b\sum_{i} \hat\rho^{(i)}_{-\bm q}(1-\mathcal{P}_b)\hat\rho^{(i)}_{\bm q} \, \mathcal{P}_b,
\end{align}
where in the last step we used that $\rho^{(i)}_{\bm q}$ only has components in the $i$th sector of the $N$-particle product space, such that $\mathcal{P}_b\rho^{(i)}_{-\bm q}(1-\mathcal{P}_b)$ vanishes everywhere except in the $i$th sector, and a similar argument holds for $(1-\mathcal{P}_b)\hat\rho^{(j)}_{\bm q} \, \mathcal{P}_b$.
Remarkably, the operator $\hat{S}^r_{\bm{q}b}$ is thus a single-particle operator, and will have a simple quadratic form in second quantization.

\subsection{Second quantization operators in the Bloch basis}

For a periodic system, we can use a Bloch basis of states $\ket{\psi_{\bm{k}b}}$ with crystal momentum $\bm{k}$ and band index $b$.
The wavefunctions are modulated plane waves $\psi_{\bm{k}b}(\bm{r})=\braket{\bm{r}|\psi_{\bm{k}b}}=e^{i \bm{k}\cdot\bm{r}} u_{\bm{k}b}(\bm{r})$, where $u_{\bm{k}b}(\bm{r})$ is periodic.
In terms of Fourier coefficients, $\psi_{\bm{k}b}(\bm{r})=e^{i\bm{k}\cdot\bm{r}} \sum_{\bm{G}} e^{i\bm{G}\cdot\bm{r}} u^{\bm{G}}_{\bm{k}b}$.
For practical purposes, it becomes natural to group the Fourier components into a vector representation $\ket{u_{\bm{k}b}}\equiv(u_{\bm{k}b}^{\bm{G}_1}, u_{\bm{k}b}^{\bm{G}_2}, \dots)$ given a fixed basis of reciprocal lattice vectors $\left\lbrace\bm{G}_1,\bm{G}_2,\dots\right\rbrace$.

The inner products between the periodic wavefunctions (integrals in real-space coordinates) simplify to $\braket{u_{\bm{k}b}|u_{\bm{k}'b'}}=\sum_{\bm{G}} u^{\bm{G}\,*}_{\bm{k}b}\,u^{\bm{G}}_{\bm{k}'b'}$.
The Bloch states are unique only in the first Brillouin zone, and we can set $\ket{\psi_{\bm{k}-\bm{G} b}}\equiv\ket{\psi_{\bm{k} b}}$, such that $\ket{u_{\bm{k}-\bm{G}b}}=e^{i\bm{G}\cdot\hat{\bm{r}}}\ket{u_{\bm{k}b}}$.
Exploiting this gauge choice allows to map any inner product of momenta in extended momentum space to the first Brillouin zone:
\begin{align}\label{eq:bloch_innerproduct}
	\braket{u_{\bm{k}b}|u_{\bm{k}'b'}} = \braket{u_{\lfloor\bm{k}\rfloor b}|e^{i(\bm{g}-\bm{g}')\cdot\hat{\bm{r}}}|u_{\lfloor\bm{k}'\rfloor b'}}
	=
	\sum_{\bm{G}} \left(u^{\bm{G}-\bm{g}}_{\lfloor\bm{k}\rfloor b}\right)^* u^{\bm{G}-\bm{g}'}_{\lfloor\bm{k}'\rfloor b'},
\end{align}
where  $\lfloor\bm{k}\rfloor=\bm{k}-\bm{g}$ is reduced to the first Brillouin zone through a suitable shift by reciprocal lattice vector $\bm{g}$.

In second quantized form, the Fock-space particle density operator $\hat{\rho}(\bm{r})=\hat{\psi}^{\dagger}(\bm{r})\hat{\psi}(\bm{r})$ is given by the field operator $\hat{\psi}^\dagger(\bm{r})=\sum_{\bm{k}\in\mathrm{BZ},b} \psi_{\bm{k}b}^*(\bm{r}) \, a^\dagger_{\bm{k}b}$ for a periodic system, using the Bloch wavefunctions $\psi_{\bm{k}b}$.
Like with the eigenstates, we can set $a^\dagger_{\bm{k}-\bm{G}b}=a^\dagger_{\bm{k}b}$.
In this representation, we find
\begin{align}
	\hat{\rho}_{\bm{q}}
	 & =  \sum_{\bm{k}\in\mathrm{BZ},bb'} \braket{\psi_{\bm{k}-\bm{q}b}|e^{-i\bm{q}\cdot\hat{\bm{r}}}|\psi_{\bm{k}b'}} a^\dagger_{\bm{k}-\bm{q}b} a^{\pdagger}_{\bm{k}b'} \nonumber \\
	 & = \sum_{\bm{k}\in\mathrm{BZ},bb'} \braket{u_{\bm{k}-\bm{q}b}|u_{\bm{k}b'}} a^\dagger_{\bm{k}-\bm{q}b} a^{\pdagger}_{\bm{k}b'}
	=\sum_{\bm{k}\in\mathrm{BZ},bb'} \braket{u_{\lfloor\bm{k}-\bm{q}\rfloor b}|e^{i\bm{g}\cdot\hat{\bm{r}}}|u_{\bm{k}b'}} a^\dagger_{\lfloor\bm{k}-\bm{q}\rfloor b} a^{\pdagger}_{\bm{k}b'}.
\end{align}
From here on, we will implicitly assume that summations over Bloch momenta are restricted to the first Brillouin zone.

As a reminder, we can introduce operators from their single-particle counterparts as $ \hat A(\bm{r},\bm{r'}) = \braket{\bm{r}|A|\bm{r'}} \hat\psi^\dagger(\bm{r})\hat\psi(\bm{r'}), $ and then proceed to calculate many-body expectation values $\braket{A(\bm{r},\bm{r'})}$.
For example, for the density operator, this yields
\begin{align}
	\rho^{(1)}(\bm{r},\bm{r'})
	= \braket{\hat\psi^\dagger(\bm{r})\hat\psi(\bm{r'})}
	= \sum_{\bm{k}b,\bm{k}'b'} \braket{\psi_{\bm{k}b}|\bm{r}} \braket{\bm{r}'|\psi_{\bm{k}'b'}} \braket{a^\dagger_{\bm{k}b}a^{\pdagger}_{\bm{k}'b'}}
	= n \, \delta(\bm{r}-\bm{r}'),
\end{align}
where $n=N/A$ is the particle density per area.
Similarly, for the single-particle quantity $\rho_{\bm{k}b}=\ket{\psi_{\bm{k}b}}\bra{\psi_{\bm{k}b}} \braket{a^\dagger_{\bm{k}b}a^{\pdagger}_{\bm{k}b}}$, we can introduce the density operator
$
	\hat\rho_{\bm{k}b}(\bm{r},\bm{r'})
	= \frac{1}{N} \braket{\bm{r}|\rho_{\bm{k}b}|\bm{r'}} \hat\psi^\dagger(\bm{r})\hat\psi(\bm{r'})
$
such that
\begin{align}
	\int d^2 r d^2 r' \braket{\hat\rho_{\bm{k}b}(\bm{r},\bm{r'})}
	= \int d^2 r d^2 r' \delta(\bm{r}-\bm{r'})\,|\psi_{\bm{k}b}(\bm{r})|^2 \, \braket{a^\dagger_{\bm{k}b}a^{\pdagger}_{\bm{k}b}}
	= \braket{a^\dagger_{\bm{k}b}a^{\pdagger}_{\bm{k}b}}
\end{align}

The structure factors $\hat{S}_{\bm{q}}$, $\hat{S}_{\bm{q}b}$ and $\hat{\mathcal{S}}_{\bm{q}b}$ are fundamentally two-particle operators.
However, the difference $\hat{S}^r_{\bm{q}b}$ is a single-particle operator: From \cref{eq:S_diff} and using $\hat\rho^{(i)}_{\bm{q}} = \sum_{\bm{k}bb'} \braket{u_{\bm{k}-\bm{q}b}|u_{\bm{k}b'}} \ket{\psi_{\bm{k}-\bm{q}b}}\bra{\psi_{\bm{k}b'}}$, we immediately find a simple single-particle representation by evaluating matrix elements $\braket{\psi_{\bm{k}b} |\hat{S}^r_{\bm{q}b} | \psi_{\bm{k}'b} }$.
This results in
\begin{align}
	\hat{S}^r_{\bm{q}b} =
	\sum_{\bm k} \Big[ 1 - |\braket{ u_{\bm{k}b} | u_{\bm{k}-\bm{q}b} } |^2  \Big]
	\,
	a^{\dagger}_{\bm{k} b} a^{\pdagger}_{\bm{k}b}.
\end{align}

As discussed in the main text, in the long-wavelength limit, we can use the Fubini-Study metric tensor $ |\braket{u_{\bm{k}b} | u_{\bm{k}+d\bm{k}b}}|^2 = 1 - g^b_{\mu\nu}(\bm{k}) \, d k_{\mu} d k_{\nu}, $ to write $ \lim_{q\to 0} \hat{S}^r_{\bm{q}b} = q_{\mu} q_{\nu} \sum_{\bm k} g^b_{\mu\nu}(\bm{k}) \, a^{\dagger}_{\bm{k} b} a^{\pdagger}_{\bm{k}b}.
$

\subsection{Expectation values: Fermi liquid states}

To this point, we have discussed operator identities.
Let us now assume that our ground state $\ket{\Psi_0}$ is a Fermi liquid, i.e., it can be expressed through long-lived quasi-particles that obey Fermi-Dirac statistics.
We can then work with the single-particle density matrix
\begin{align}
	G_{\bm{k}} = \sum_{b} \ket{u_{\bm{k}b}}\bra{u_{\bm{k}b}} n_F(\epsilon_{\bm{k}b}-\mu), \end{align} where $n_F$ is the Fermi-Dirac distribution, $\mu$ is the chemical potential.
We can evaluate the  expectation values:
\begin{align}\label{eq:structurefactors_fermiliquic1}
	S_{\bm{q}} = \frac{1}{N} \braket{\hat\rho_{-\bm q} \hat\rho_{\bm q}} & = \frac{1}{N} \sum_{\bm{k}}\tr G_{\bm{k}} -\frac{1}{N} \sum_{\bm{k},\bm{k}'} \tr\left[G_{\bm{k}}\left(\delta_{\bm{q},0}\tr[G_{\bm{k'}}] - \delta_{\bm{k}-\bm{k'},\bm{q}} G_{\bm{k'}} \right)\right], \\ \label{eq:structurefactors_fermiliquic2} S_{\bm{q}b} & = \frac{1}{N} \sum_{\bm{k}}\tr \left[\mathcal{P}_bG_{\bm{k}}\right] -\frac{1}{N} \sum_{\bm{k},\bm{k}'} \tr\left[\mathcal{P}_b G_{\bm{k}}\left(\delta_{\bm{q},0}\tr[[\mathcal{P}_b G_{\bm{k'}}] - \delta_{\bm{k}-\bm{k'},\bm{q}} G_{\bm{k'}} \right)\right], \\ \label{eq:structurefactors_fermiliquic3} \mathcal{S}_{\bm{q}b} & = \frac{1}{N} \sum_{\bm{k}}\tr \left[\mathcal{P}_bG_{\bm{k}}\right] -\frac{1}{N} \sum_{\bm{k},\bm{k}'} \tr\left[\mathcal{P}_b G_{\bm{k}}\left(\delta_{\bm{q},0}\tr[\mathcal{P}_b G_{\bm{k'}}] - \delta_{\bm{k}-\bm{k'},\bm{q}} \mathcal{P}_b G_{\bm{k'}} \right)\right], \\ \label{eq:structurefactors_fermiliquic4} S^r_{\bm{q}b} & = \frac{1}{N} \sum_{\bm{k}} \tr\left[\mathcal{P}_b G_{\bm{k}} \left(1- \mathcal{P}_b \right) G_{\bm{k}-\bm{q}} \right] .
\end{align}

It is a now straightforward matter to write the structure factors in
\cref{eq:structurefactors_fermiliquic1,eq:structurefactors_fermiliquic2,eq:structurefactors_fermiliquic3,eq:structurefactors_fermiliquic4} in terms of Fourier coefficients, for example
\begin{align}
	S^r_{\bm{q}b} = \frac{1}{N} \sum_{\bm{k}} \left[1-|\braket{u_{\bm{k}-\bm{q}b}|u_{\bm{k}b}}|^2\right] \, n_F(\epsilon_{\bm{k}b}-\mu).
\end{align}

\section{Zero-temperature expectation values and the Markov chain Monte Carlo method}

\paragraph{General.} In the study of quantum many-body systems, a fundamental object is the $N$-particle wavefunction
\begin{equation}
	\Psi(\bm{r_1},\ldots,\bm{r_N}) = \braket{\bm{r}_1,\dots,\bm{r}_N|\Psi},
\end{equation}
of a state $\ket{\Psi}$ for a given set of coordinates \( \bm{R}\equiv(\bm{r_1},\ldots,\bm{r_N}) \).
The associated probability density is $w(\bm{R}) = |\Psi(\bm{R})|^2$ and describes the likelihood of observing the system in a particular configuration \( \bm{R} \).
The expectation value of a local observable $\hat A$ (that is, $\braket{\bm{R}|\hat A|\bm{R'}}=\delta(\bm{R}-\bm{R'}) A(\bm{R})$) in state $\ket{\Psi}$ can be written as
\begin{equation}\label{eq:manybody_expval}
	\braket{\hat{A}}_{\Psi}
	= \braket{\Psi|\hat{A}|\Psi}
	= \int d^{DN}
	R \, |\Psi(\bm{R})|^2 A(\bm{R}) \equiv \int d^{DN}R \, \frac{e^{-S(\bm{R})}}{Z} A(\bm{R}) \approx \frac{1}{M} \sum_{r=1}^M A(\bm{R}^{(r)}), \end{equation} where, in the last step, we approximate the $dN$-dimensional integral using Monte-Carlo importance sampling: $\mathcal{M}=\left\lbrace\bm{R}^{(1)},\dots,\bm{R}^{(M)}\right\rbrace$ is a set of $M$ samples drawn from the probability distribution $w(\bm{R})=|\Psi(\bm{R})|^2$.
In practice, these samples are efficiently generated from a Markov chain using the Metropolis-Hastings algorithm.
We emphasize that the zero-temperature quantum expectation value $\braket{\hat{A}}_{\Psi}$ thus maps to a finite-temperature classical gas expectation value with classical action $S(\bm{R})\equiv-\log |\Psi(\bm{R})|^2 + \text{const}$.

\paragraph{Laughlin wavefunction.} As motivated in the main text, the Laughlin many-body wavefunction is of fundamental importance in the context of fractional Quantum Hall physics, and is given by
\begin{align}
	\Psi_m(\bm{r}_1,\dots,\bm{r}_N) = \prod_{i<j}^{N}(z_i\!-\!z_j)^m\prod_{k=1}^{N}e^{-\frac{|z_k|^2}{4\ell^2}-u(\bm{r}_k)},
\end{align}
where $m$ is an odd integer,
$z_k=x_k+iy_k$ is position expressed as a complex number, $\ell$ is the magnetic length such that we have one magnetic flux per unit cell area $A_0=2\pi \ell^2$, and $u(\bm{r})$ is periodic under lattice translations and averages to zero.
For this case, it is a straightforward exercise to show the classical action $S[\rho_{\bm{R}}]$ introduced in \cref{eq:manybody_expval} is (up to const.)
\begin{align}\label{eq:plasma_action_real}
	S[\rho_{\bm{R}}] = - \sum_{\bm{r}} \phi_{\text{b}}(\bm{r}) \rho_{\bm{R}}(\bm{r}) + \frac{m}{2} \sum_{\bm{r},\bm{r}'} \rho_{\bm{R}}(\bm{r}) V(\bm{r}-\bm{r}') \rho_{\bm{R}}(\bm{r}') , &  & \text{with } &  & \phi_{\text{b}}(\bm{r})= \sum_{\bm{r'}} \, V(\bm{r}-\bm{r}') \, \rho_{\text{b}}(\bm{r}'),\end{align} where we defined the particle density $\rho_{\bm{R}}(\bm{r}) = \sum_i \delta_{\bm{r},\bm{r}_i}$, $V(\bm{r})\equiv-2\ln|\bm{r}|$ and $\phi_{\text{b}}(\bm{r})$ is the attractive 2D Coulomb potential produced by the background charge $\rho_{\text{b}}(\bm{r})= \Delta A/(2\pi\ell^2) + \Delta A \nabla^2 u(\bm{r})/2\pi$ with $\Delta A = A/N_p$ and $N_p$ is the spatial discretization.
Remarkably, the action $S[\rho_{\bm{R}}]$ in this case is the same as that of an interacting classical gas of repulsive particles in a positive background, a mapping known as plasma analogy.

We can restate $S[\rho_{\bm{R}}]$ using Fourier representations (i.e.,
$\rho_{\bm{R}}(\bm{r}) = N_p^{-1} \sum_{\bm{k}} \sum_i e^{i \bm{k}\cdot(\bm{r}-\bm{r}_i)} $
and
$\rho_{\text{b}}(\bm{r}) = N_p^{-1}\cdot \sum_{\bm{G}} e^{i\bm{G}\cdot\bm{r}}\rho_{\text{b}\bm{G}}$ with $\rho_{\text{b}\bm{0}}=A/( 2\pi\ell^2)$), to find
\begin{align}\label{eq:plasmaanalogy_action_k}
	S[\rho_{\bm{R}}] & = -\sum_{\bm{k}} \rho_{\bm{R}}(-\bm{k}) \phi_{\mathrm{b}}(\bm{k})+\frac{m}{2 A} \sum_{\bm{k}} \rho_{\bm{R}}(-\bm{k}) V(\bm{k}) \rho_{\bm{R}}(\bm{k})
\end{align}
with $\phi_{\mathrm{b}}(\bm{k})=\frac{1}{A} V(\bm{k}) \rho_{\mathrm{b} \bm{k}}$, which we can further rewrite to
\begin{align*}
	S[\rho_{\bm{R}}] & = \frac{1}{2A} \sum_{\bm{k}\ne 0} \frac{4\pi m}{|\bm{k}|^2} \; |\rho_{\bm{R}}(\bm{k})-\frac{\rho_{\text{b}\bm{k}}}{m}|^2 + \underbrace{\frac{1}{2A} V(\bm{k}\to 0)\, |N-\frac{\rho_{\text{b}\bm{k}=\bm{0}}}{m}|^2}_{\Rightarrow \; n = \frac{N}{A} = \frac{1}{m} \frac{1}{2\pi \ell^2}=\frac{1}{m} \frac{1}{A_0}} + \text{ const}.
	=
	\frac{1}{2A} \sum_{\bm{k}\ne 0} \frac{4\pi m}{|\bm{k}|^2} \; |\rho_{\bm{R}}(\bm{k})-\frac{\rho_{\text{b}\bm{k}}}{m}|^2
	+ \text{ const}.
\end{align*}
Here, we absorbed the self-interaction terms into an overall normalization constant, as they do not depend on the configuration $\bm{R}$.
Furthermore, from the second term, we see that the wavefunction only has a well-defined norm in the thermodynamic limit ($N\to\infty$) if $n = N/A = 1/(m 2\pi \ell^2)=1/(mA_0)$.
This puts a fundamental physical constraint on the wavefunction.

In practice, to carry out Monte-Carlo sampling, we consider a system area that contains $N_{\phi}=N_1\, N_2$ unit cells, i.e., $A=N_{\phi} A_0$.
The unit cell area is $A_0=|\bm{a}_1\times \bm{a}_2|$, where lattice vectors $\bm{a}_j$ do not need to be orthogonal, and $\bm{b}_j$ are corresponding reciprocal lattice vectors.
We discretize spatial coordinates such that each unit cell contains $n_1\times n_2$ points.
We parameterize the real-space grid as $\bm{r}_{ij}=i/n_1 \bm{a}_1 + j/n_2 \bm{a}_2$ with $i=0,\dots,n_1 N_1-1$ and $j=0,\dots,n_2 N_2-1$.
Similarly, in reciprocal space, we get the grid $\bm{k}_{ij} = -\left((1/2) n_1 + i /N_1\right) \bm{b}_1 + \left((1/2) n_2 + j /N_2\right) \bm{b}_2$.
We fix that the system is periodic at the boundaries of area $A$.
Notably, we can then represent the real-space particle density through matrices $\rho_{\bm{R}}^{ij}\equiv \rho_{\bm{R}}(\bm{r}_{ij})$ that contain $0$ and $1$ entries.
We obtain the corresponding reciprocal space density $\hat{\rho}^{ij}_{\bm{R}}\equiv\rho_{\bm{k}_{ij}}(\bm{R})$ through fast Fourier transform.
The density constraint $n=1/(mA_0)$ implies the total particle number $N=N_{\phi}/m$ in our system area.

\paragraph{Linear response.}
We are interested in the case where the effective magnetic field modulation (that enters as $u(\bm{r})$ in the Laughlin wave function) is several times smaller than the homogeneous background.
As we discussed and numerically confirmed in the main text, this justifies the following linear response approximation.
The response coefficient is the derivative of \cref{eq:manybody_expval} with respect to a coupling between the observable and the perturbing field.
In the case of  \cref{eq:plasmaanalogy_action_k}, we find
\begin{align}
	\left.\frac{\partial \rho(\bm{G})}{\partial \phi_b(\bm{G})}\right|_{\phi_b(\bm{G})=0}=\left.\frac{\partial\left\langle\rho_{\bm{R}}(\bm{G})\right\rangle}{\partial \phi_b(\bm{G})}\right|_{\phi_b(\bm{G})=0}=-\left.
	N S(\bm{G})\right|_{\phi_b(\bm{G})=0},\end{align} which agrees with the fact that the structure factor is the density-density response.
With minor manipulations, we thus arrive at the linear response relation
\begin{align}
	\rho(\bm{G})=-\left.\bar{n} \; V(\bm{G}) S(\bm{G})\right|_{\rho_{\mathrm{b}
		G}=0} \rho_{\mathrm{b} \bm{G}} = -\left.
	\frac{1}{A_0} \; V(\bm{G}) S(\bm{G})\right|_{\rho_{\mathrm{b}
		G}=0} \frac{\rho_{\mathrm{b}\bm{G}}}{m} .
\end{align}

\end{document}